\documentclass[useAMS,usenatbib]{mn2e}
\usepackage{epsfig}
\usepackage{color}
\usepackage{subfigure}
\usepackage{amsmath}
\usepackage{amssymb}

\def\lsim{\mathrel{\rlap{\lower3.5pt\hbox{\hskip0.5pt$\sim$}}
    \raise0.5pt\hbox{$<$}}}
\def\gsim{~\rlap{$>$}{\lower 1.0ex\hbox{$\sim$}}}

\newcommand{\goodgap}{\hspace{\subfigtopskip} \hspace{\subfigbottomskip}}

\title[Cladistics and GRBs]{Hints for families of GRBs improving the Hubble diagram}
\author[V.F. Cardone \& D. Fraix\,-\,Burnet]{Vincenzo F. Cardone$^{1}$\thanks{Corresponding author\,: {\tt winnyenodrac@gmail.com}}, Didier Fraix\,-\,Burnet$^2$ \\
$^1$I.N.A.F.\,-\,Osservatorio Astronomico di Roma, via Frascati 33, 00040 - Monte Porzio Catone (Roma), Italy \\
$^2$Institut de Plan\'etologie et d'\,Astrophysique de Grenoble, Universit\'e Joseph Fourier - Grenoble 1/CNRS, 38041\,-\,Grenoble, France}

\date{Accepted xxx, Received yyy, in original form zzz}

\begin{document}

\maketitle

\begin{abstract}

As soon as their extragalactic origins were established, the hope to make Gamma\,-\,Ray Bursts (GRBs) standardizeable candles to probe the very high\,-\,$z$ universe has opened the search for scaling relations between redshift independent observable quantities and distance dependent ones. Although some remarkable success has been achieved, the empirical correlations thus found are still affected by a significant intrinsic scatter which downgrades the precision in the inferred GRBs Hubble diagram. We investigate here whether this scatter may come from fitting together objects belonging to intrinsically different classes. To this end, we rely on a cladistics analysis to partition GRBs in homogenous families according to their rest frame properties. Although the poor statistics prevent us from drawing a definitive answer, we find that both the intrinsic scatter and the coefficients of the $E_{peak}$\,-\,$E_{iso}$ and $E_{peak}$\,-\,$L$ correlations significantly change depending on which subsample is fitted. It
turns out that the fit to the full sample leads to a scaling relation which approximately follows the diagonal of the region delimited by the fits to each homogenous class. We therefore argue that a preliminary identification of the class a GRB belongs to is necessary in order to select the right scaling relation to be used in order to not bias the distance determination and hence the Hubble diagram.

\end{abstract}

\begin{keywords}
Gamma Rays\,: bursts -- Cosmology\,: distance scale -- methods: statistical
\end{keywords}

\section{Introduction}

The astonishingly high energy of their almost instantenous $\gamma$\,-\,ray emission makes Gamma Ray Bursts (GRBs) one of the most (if not the most) energetic astrophysical phenomena. Moreover, thanks to their high luminosity, GRBs are visible up to very high redshift thus attracting a lot of interest for a potential application to trace the Hubble diagram deep into the matter dominated era. To this end, a lot of work has been devoted to finding and characterizing scaling relations among observables redshift independent quantities (e.g., the time duration) and distance dependent ones (such as the isotropic luminosity and energy) so that the Hubble diagram can be inferred. Although significant steps forward have been done with the discovery of different 2D correlations, the remarkable scatter affecting them and the uncertainties on their physical interpretation makes this issue still open and strongly debated.

It is worth noting that all the 2D relations investigated so far have been first discovered examining a relatively small number of objects. It is therefore not surprising that the main aim of later analyses has been to add GRBs to the fitted samples in order to both strengthen the significance of the correlation and ameliorate the statistics. The recent availability of the Swift and Fermi satellites have, however, changed this situation greatly increasing the total number of GRBs and hence the ones usable for the analysis of different scaling laws of interest. It is therefore possible now to deepen the investigation by separating GRBs in different classes according to their observable properties.

The importance of a correct partitioning of astronomical objects in homogenous groups can be easily understood considering the textbook example of the Cepheids variables and the estimate of the Hubble constant $H_0$. Indeed, the first $H_0$ determinations based on a mixed sample of Cepheids and RR Lyrae variables were grossly wrong because of the use of the same period\,-\,luminosity relation for two radically different stellar objects. Another instructive example is represented by Supernovae. Indeed, no scaling relations can be found if one considers the full sample, but this becomes possible if one separates the Type Ia class SNe from the other ones. Motivated by these examples, we investigate here the impact of partitioning GRBs in homogenous classes on the slope and scatter of the popular Amati relation.

Actually, defining homogenous classes of GRBs is far from trivial. As in every classification analysis, one has first to decide which are the criteria to use. Put in other words, one should decide how to separate a multidimensional parameter space in disjoint regions. Two alternative and somewhat complementary choices are possible. On one hand, one can start from a (sometimes a priori) theoretical interpretation of the classification. For instance, it was the vague (and later found to be wrong) idea that spiral galaxies represent different possible evolutionary stages of elliptical ones which leads Hubble to formulate its tuning fork diagram. On the contrary, one can start from some observable properties (such as, e.g., the period\,-\,luminosity relations for variable stars or the presence or lack of some chemical elements in the Supernova spectrum) to work out a classification. Ideally, the two approaches should converge with theory leading to classes with well distinct observational properties and
observations suggesting theoretical interpretation of the physical phenomena underlying the empirical partitioning. Unfortunately, notwithstanding the large number of GRBs nowadays available, classification of GRBs is still in its infancy. Usually, one adopts a very rough classification dividing GRBs in short (with $T_{90} \le 2 \ {\rm sec}$) and long $(T_{90} > 2 \ {\rm sec})$ bursts where $T_{90}$ is a typical time duration (better defined later). Indeed, the $T_{90}$ distribution clearly shows two distinct peaks thus suggesting that different physical phenomena are taking place for the two families. Theoretically, long bursts are associated to supernova explosion so that $T_{90}$ is determined by the time needed for the dynamical collapse of the exploding star. As such, one expects to detect long bursts in starburst regions and to find a supernova when following up the GRBs in optical bands. In contrast, the merger of two neutron stars may be the origin of short bursts since such a merging takes place on
short timescale. There are nevertheless other possible physical origins for GRBs such as the merger of a neutron star with a white dwarf that should lead to an intermediate class. Moreover, it is not unusual to find long bursts with no associated supernova thus suggesting that a clear cut separation in only two families is too restrictive and possibly misleading.

A pioneering step forward along the road to classify GRBs in more than one family has been taken by \cite{Chat07}. Using two different multivariate clustering techniques, namely the K\,-\,means partitioning method and the Dirichlet process of mixture modeling, applied to the BATSE GRB catalog \citep{Batse}, \cite{Chat07} found that the optimal number of classes is three in agreement with a theoretical speculation interpreting the three classes as coming out from mergers of neutron star systems, mergers between white dwarfs and neutron stars, and collapse of massive stars. It is worth noting, however, that \cite{Chat07} have used the full BATSE sample notwithstanding the availability of redshift. As a consequence, they could not correct the observed quantities to the GRB rest frame so that their classification is not based on the intrinsic GRBs properties, but only on their apparent ones. As a further step on, we revisit here the issue of partitioning GRBs improving the analysis in two aspects. First, we rely
on GRBs with known $z$ so that we can deal with rest frame quantities which are more intimately related to the GRB physics. Second, we resort to cladistics to separate GRBs in distinct families taking into account a large parameter space. Cladistics has been developed to seek for evolutionary relationship among objects so that is particularly well suited for our aims offering the possibility to work out classes populated by GRBs with homogenous properties.

The plan of the paper is as follows. Sect.\,2 gives a short introduction to cladistics also explaining why we believe this can be a valuable help in addressing the problem of GRBs classification. The GRBs sample and the parameters used as input to the analysis are described in Sect.\,3 where we  provide some qualitative interpretation of the resulting families. As an application, we examine in Sect.\,4 how the slope and the scatter of two popular scaling relations depend on the particular GRBs class used as input to the fitting procedure showing the importance of a preliminary separation of GRBs in homogenous subsamples to avoid biasing such relations. A summary of the results is finally given in Sect.\,5.

\section{Cladistics}

Cladistics belongs to the phylogenetic methods, designed to build a graph representing the evolutionary relationships between species of objects (\citealt{Felsenstein2003,Makarenkov2006}). Each node of the graph indicates a transmission with modification mechanism that creates two or more species inheriting from a common ancestor.

Cladistics is a non\,-\,parametric parameter\,-\,based method also called the maximum parsimony method  \citep{semple2003}. There is no assumption about the metrics of the parameter space.  The principle of cladistics is relatively simple: two (or more) objects are related if they share a common history, that is they possess properties inherited from a common ancestor. The parameters (called {\it characters}) are traits, descriptors, observables, or properties, which can be assigned at least two states characterising the evolutionary stage of the objects for that parameter. The maximum parsimony algorithm looks for the simplest arrangement of objects on a bifurcating tree. The complexity of the arrangement is measured by the total number of  {\it steps} (i.e. changes in all parameter states) along the tree. The simplest tree supposedly depicts the simplest evolutionary scenario.

Cladistics has been mainly developed to build phylogenies of living organisms and is well adapted to the diversification by replication. More generally, cladistics can be used when diversity is generated through some transmission\,-\,with\,-\,modification mechanism, ideally presenting a branching pattern.

Cladistics can take uncertainties into account. This is an invaluable capability that is rather rare among clustering methods. The implementation is very simple since it suffices for each parameter to be given a range of values instead of a single value. The algorithm then evaluates the different possibilities allowed by the range of values and select among them the one that provides the most parsimonious tree. In the same way, undocumented parameters can be included, and the most parsimonious diversification scenario provides a prediction for the unknown values.

The idea behind astrocladistics, initially implemented for galaxies \citep{FCD06,jc1,jc2} is that the evolution of astrophysical objects is most often a transformation, which is nothing else than a transmission with modification. For galaxies this occurs when they are transformed through assembling, internal evolution, interaction, merger or stripping
\cite{FCD06,jc1,jc2,DFB09}. For each transformation event, stars, gas and dust are transmitted to the new object with
some modification of their properties.

The globular clusters can be seen as simple forms of galaxies, without interaction. Their properties and their evolution depend mainly on the environment in which they form. Since this environment evolves itself (evolution of the Universe and the dynamical environment of the parent galaxy), the basic properties of different clusters are related to each other by some evolutionary pattern. In particular, the dust and gas from which the stars of the globular clusters form have been "polluted" (enriched in heavy elements) by more ancient stars, being field stars or belonging to other globular clusters. This results in a kind of transmission with modification process, which justifies a priori the use of cladistics on globular clusters, as confirmed in \citet{FDC09}.

The specific concept behind cladistics that justifies its application to other astrophysical entities such as GRBs may be more easily understood in the case of stars (Fraix-Burnet \& Thuillard, A\&A, submitted). A single star with a given mass and metallicity evolves along a well\,-\,defined trajectory usually represented in the Herztsprung\,-\,Russell diagram that plots the luminosity versus the temperature. All stars with the same mass and metallicity follow the same evolutionary path, and we can see these stars as belonging to a same lineage or family. Several such lineages create a graph, with branches connecting at the Main Sequence. It is thus quite natural to use classification methods that are designed to build graphs. Cladistics, being non-parametric and parameter based, is the most general and powerful of them thus emerging as the most suitable choice.

We may expect GRBs, as well as many other astrophysical objects, to obey more or less the same rule: from a given state, they evolve along a path that is characterized by the properties of the lineage at the initial state. Hence, provided we have enough objects to sample the evolutionary paths of the lineages, and assuming the available parameters (observables) contain enough information to distinguish a few lineages, cladistics seems to be an adequate tool to use.

The important consequence for our study of GRBs is that since the objects within each family are by construction, homologous, it is reasonable to expect the dispersion in the scaling relations to be reduced.

In the present study, we have performed a cladistic analysis in the same way as done in several previous papers (see, e.g., \citealt{Fraix2012})\,: each parameter is discretized into 30 equal-width bins, which play the role of discrete evolutionary states. We have taken into account the measurement errors available for $\log{E_{iso}}$, $\log{L_{iso}}$\ and $HR$ as ranges of possible values (see above). The search for the maximum parsimony trees are performed using the heuristic algorithm implemented in the PAUP*4.0b10 \citep{paup} package, with the Ratchet method \citep{ratchet}.

\section{Partitioning GRBs}

The cladistics approach described above allows us to separate GRBs in classes according to a set of properties used to generate the most parsimonious \textbf{tree}. The members of a class share similar properties (likely related to their formation and evolutionary process) thus representing a homogenous population. Partitioning GRBs using cladistics therefore offers the possibility to both infer constraints on the GRBs physics and investigating whether the slope and the scatter of empirical 2D scaling relations depend on the GRB status. To this end, we here first briefly describe the input data and then show the results of the cladistics analysis.

\subsection{The data}

In order to avoid any possible systematics related to merging data from different sources, we only rely on GRBs observed by the {\it Swift} satellite (\cite{G04}) retrieving their data from the online archive\footnote{{\tt http://swift.gsfc.nasa.gov/docs/swift/archive/grb\_table/}}. Since we need rest frame and intrinsic quantities, we search for GRBs with measured $z$ which reduces the full sample to 216 objects. We then select GRBs with measured values of the quantities listed below.

\begin{itemize}

\item[-]{$T_{90}$\,: the time (in $s$) over which the burst emits from $5$ to $95\%$ of its total measured counts;}

\item[-]{${\cal{S}}$\,: the fluence (in $10^{-7} \ {\rm erg/cm^2}$) measured over the energy range $(15, 150) \ {\rm keV}$;}

\item[-]{${\cal{P}}$\,: the peak phothon flux (in ${\rm ph/cm^2}/s$) over the same energy range;}

\item[-]{${\cal{F}}_X$\,: the XRT early flux (in $10^{-11} \ {\rm erg/cm^2}/s$) measured over the energy range $(0.3, 10) \ {\rm keV}$;}

\item[-]{$\gamma$\,: the spectral index, i.e., the slope of the X\,-\,ray power spectrum approximated as a power\,-\,law.}

\item[-]{${\cal{N}}_{HI}$\,: the column density of the circumburst material.}

\end{itemize}
Not all of these quantities can be directly used as input to the cladistics analysis since they are observable rather than intrinsic GRB properties. To this end, we will consider the rest frame time duration

\begin{equation}
\tau_{90} = T_{90}/(1 + z)
\label{eq: deft90rf}
\end{equation}
instead of $T_{90}$ as time related quantity\footnote{Whenever possible, we replace the $T_{90}$ value provided on the Swift online archive with the one given in \cite{Saka11} since this latter catalog presents refined estimate of this quantity.}. We then replace the observed fluence ${\cal{S}}$ with the isotropic emitted energy evaluated as (\cite{S07,CCD09})

\begin{equation}
E_{iso} = 4 \pi d_L^2(z) {\cal{S}}_{bolo} (1 + z)^{-1}
\label{eq: defeiso}
\end{equation}
where $d_L(z)$ the luminosity distance\footnote{We adopt a flat $\Lambda$CDM model with $(\Omega_M, h) = (0.266, 0.710)$.} and ${\cal{S}}_{bolo}$ the bolometric fluence estimated as

\begin{equation}
{\cal{S}}_{bolo} = {\cal{S}} \times \frac{\int_{1/(1 + z)}^{10^4/(1 + z)}{E \Phi(E) dE}}{\int_{E_{min}}^{E_{max}}{E \Phi(E) dE}}
\label{eq: defsbolo}
\end{equation}
with $\Phi(E)$ the GRB spectrum and $(E_{min}, E_{max}) = (15, 150) \ {\rm keV}$. Similarly, we do not use the observed peak flux ${\cal{P}}$, but the isotropic luminosity given by (\cite{S07,CCD09})

\begin{equation}
L_{iso} = 4 \pi d_L^2(z) {\cal{P}}_{bolo}
\label{eq: deflum}
\end{equation}
where the bolometric peak flux reads

\begin{equation}
{\cal{P}}_{bolo} = {\cal{P}} \times \frac{\int_{1/(1 + z)}^{10^4/(1 + z)}{E \Phi(E) dE}}{\int_{E_{min}}^{E_{max}}{\Phi(E) dE}}\ .
\label{eq: defpbolo}
\end{equation}
In order to evaluate $(E_{iso}, L_{iso})$, we need to model the GRB spectrum $\Phi(E)$ over the $(E_{min}, E_{max})$ range. As a first approximation, one could set $\Phi(E) \propto E^{-\beta}$ and rely on the $\beta$ values reported in the online Swift archive. Actually, most GRB spectra are better approximated by the \cite{B93} function

\begin{equation}
\Phi(E) = \left \{
\begin{array}{ll}
\displaystyle{A E^{\alpha} \exp{\left [- \frac{(2 + \alpha) E}{E_{pk}} \right ]}} & \displaystyle{E \le
\left ( \frac{\alpha - \beta}{2 + \alpha} \right ) E} \\
� & � \\
\displaystyle{B E^{\beta}} & {\rm otherwise} \\
\end{array}
\right . \ ,
\label{eq: bandphi}
\end{equation}
with $(E_{pk}, \alpha, \beta)$ determined from the fit to the observed spectrum and $(A, B)$ normalization constants. Depending on the $E_{pk}$ value and the data quality, it is not always possible to determine $(E_{pk}, \alpha, \beta)$ so that (following, e.g., \citealt{S07,A13}) we set $\beta = -2.3$ if this latter is not available. We match our GRBs list with the GRB spectral catalog compiled by L. Amati (private communication, see also \citealt{A13}) thus finding spectral parameters for 90 out of 216 objects. For the remaining ones, we follow the Swift archive assuming that a power\,-\,law approximation is a reliable description of the spectrum\footnote{Note that we have also searched the literature checking that this is indeed the case for GRBs with spectra observed by other instruments.} and fixing the slope $\beta$ to the online value.

Since the spectrum is likely related to the GRB physics, it is interesting to include in the cladistics analysis a quantity related to its profile. This is provided by the {\it hardness ratio} which we compute as\,:

\begin{equation}
HR = \frac{\int_{100 \ {\rm keV}}^{100 \ {\rm keV}}{E \Phi(E) dE}}{\int_{25 \ {\rm keV}}^{50 \ {\rm keV}}{E \Phi(E) dE}}
\label{eq: hrdef}
\end{equation}
thus quantifying in which energy range the GRBs emits most of its photons.

In order to deal with rest frame quantities only, we convert the observed luminosity X\,-\,ray flux ${\cal{F}}_X$ into the corresponding luminosity $L_X$ assuming a $\Phi(E) \propto E^{\gamma}$ which represents an excellent approximation for the spectrum over the $(E_{min}, E_{max}) = (0.3, 10) \ {\rm keV}$ band.

We finally exclude from the sample three objects since they are outliers in the distribution of one of the parameters described above\footnote{The three systems are GRB060123 having an unusually large $\tau_{90}$ and (GRB061217, GRB060602A) because of the weird $\gamma$.} thus ending up with 213 GRBs with measured values of $(\tau_{90}, \log{E_{iso}}, \log{L_{iso}}, HR, \log{L_X}, \gamma, \log{{\cal{N}}_{HI}})$. These are used as input to the cladistics analysis described in the following. As a final comment, we inform the reader that errors are available only for $(\log{E_{iso}}, \log{L_{iso}}, HR)$ so that we will test the impact of these uncertainties only on our partitioning procedure.

\begin{table}
\caption{Contingency table between the nine classes found by cladistics and the six groups found by the k-medoids method. Cladistic group 0 correspond to the ten objects isolated.}
\begin{center}
\begin{tabular}{ccccccccccc}
\hline
            & \multicolumn{9}{c}{Cladistic classes}               \\
k-medoids &&&&&&&&&&\\
\cline{2-11}
groups  & 0 & 1 & 2 & 3 & 4 & 5 & 6 & 7 & 8 & 9 \\
\hline\hline
1 & 0 & 1 & 17 & 0 & 0 & 0 & 0 & 8 & 0 & 0 \\
2 & 4 & 0 & 3 & 4 & 3 & 5 & 6 & 7 & 6 & 5 \\
3 & 0 & 0 & 0 & 0 & 0 & 0 & 0 & 4 & 1 & 40 \\
4 & 0 & 0 & 0 & 6 & 0 & 0 & 0 & 0 & 4 & 8 \\
5 & 1 & 13 & 5 & 12 & 0 & 5 & 0 & 0 & 0 & 0 \\
6 & 5 & 1 & 8 & 2 & 5 & 2 & 6 & 2 & 7 & 7  \\
\hline
Total & 10 & 15 & 33 & 24 & 8 & 12 & 12 & 21 & 18 & 60  \\
\hline
\end{tabular}
\end{center}
\label{tab:contingency}
\end{table}

\begin{table*}
\caption{Median and $90\%$ CL of the GRB input parameters for the nine different classes with ${\cal{N}}_{id}$ the number of GRBs in the class $id$.}
\begin{center}
\begin{tabular}{cccccccccc}
\hline
Id & ${\cal{N}}_{id}$ & $z$ & $\tau_{90}$ & $\log{E_{iso}}$ & $\log{L_{iso}}$ & $HR$ & $\log{L_X}$ & $\gamma$ & $\log{{\cal{N}}_{HI}}$\\
\hline \hline

~ & ~ & ~ & ~ & ~ & ~ & ~ & ~ & ~ & ~\\

$1$ & 15 & $2.60_{-1.34}^{+3.70}$ & $30.12_{-19.52}^{+62.92}$ & $53.77_{-0.48}^{+0.54}$ & $61.74_{-0.57}^{+0.90}$ & $4.98_{-2.63}^{+21.2}$ & $51.20_{-1.25}^{+0.80}$ & $-1.90_{-0.13}^{+0.38}$ & $0.80_{-0.66}^{+3.09}$ \\

~ & ~ & ~ & ~ & ~ & ~ & ~ & ~ & ~ & ~\\

$2$ & 33 & $2.43_{-1.89}^{+2.87}$ & $65.18_{-42.32}^{+117.1}$ & $53.30_{-0.40}^{+0.55}$ & $61.26_{-0.75}^{+0.65}$ & $2.54_{-1.43}^{+4.29}$ & $50.87_{-2.32}^{+2.01}$ & $-2.05_{-0.46}^{+0.15}$ & $0.47_{-0.83}^{+3.34}$ \\

~ & ~ & ~ & ~ & ~ & ~ & ~ & ~ & ~ & ~\\

$3$ & 24 & $2.81_{-1.57}^{+5.19}$ & $9.19_{-8.22}^{+93.6}$ & $53.06_{-0.75}^{+0.78}$ & $61.82_{-1.06}^{+0.35}$ & $3.13_{-2.22}^{+4.14}$ & $50.92_{-1.20}^{+1.05}$ & $-1.82_{-0.39}^{+0.35}$ & $0.47_{-10.9}^{+3.22}$ \\

~ & ~ & ~ & ~ & ~ & ~ & ~ & ~ & ~ & ~\\

$4$ & 8 & $3.68_{-1.48}^{+1.92}$ & $7.28_{-5.78}^{+3.14}$ & $53.08_{-0.31}^{+0.20}$ & $61.59_{-0.41}^{+0.35}$ & $1.31_{-0.16}^{+1.29}$ & $50.54_{-0.32}^{+0.49}$ & $-2.01_{-0.19}^{+0.10}$ & $1.12_{-0.20}^{+3.55}$ \\

~ & ~ & ~ & ~ & ~ & ~ & ~ & ~ & ~ & ~\\

$5$ & 12 & $1.61_{-0.89}^{+1.26}$ & $5.03_{-4.68}^{+6.47}$ & $53.04_{-1.68}^{+0.25}$ & $61.30_{-0.77}^{+0.65}$ & $4.52_{-3.51}^{+11.4}$ & $49.78_{-0.35}^{+0.52}$ & $-1.84_{-0.10}^{+0.31}$ & $0.71_{-1.14}^{+2.45}$ \\

~ & ~ & ~ & ~ & ~ & ~ & ~ & ~ & ~ & ~\\

$6$ & 12 & $2.64_{-1.09}^{+2.47}$ & $14.74_{-8.84}^{+16.4}$ & $53.40_{-0.45}^{+0.36}$ & $61.40_{-0.43}^{+0.31}$ & $2.92_{-1.91}^{+5.21}$ & $49.94_{-2.66}^{+1.60}$ & $-2.22_{-0.28}^{+0.24}$ & $1.02_{-1.00}^{+3.54}$ \\

~ & ~ & ~ & ~ & ~ & ~ & ~ & ~ & ~ & ~\\

$7$ & 21 & $1.02_{-0.48}^{+2.35}$ & $41.67_{-27.31}^{+84.94}$ & $52.52_{-0.61}^{+0.60}$ & $60.48_{-1.19}^{+0.64}$ & $2.91_{-2.32}^{+2.59}$ & $49.64_{-1.49}^{+0.73}$ & $-2.02_{-0.32}^{+0.53}$ & $0.63_{-0.73}^{+2.93}$ \\

~ & ~ & ~ & ~ & ~ & ~ & ~ & ~ & ~ & ~\\

$8$ & 18 & $1.77_{-1.23}^{+1.58}$ & $13.43_{-13.27}^{+27.81}$ & $52.57_{-0.82}^{+0.44}$ & $60.73_{-0.79}^{+0.81}$ & $1.83_{-1.21}^{+4.03}$ & $49.84_{-0.97}^{+0.41}$ & $-2.31_{-0.44}^{+0.14}$ & $0.45_{-7.89}^{+3.45}$ \\

~ & ~ & ~ & ~ & ~ & ~ & ~ & ~ & ~ & ~\\

$9$ & 60 & $0.90_{-0.80}^{+2.30}$ & $5.50_{-5.40}^{+99.3}$ & $51.78_{-2.40}^{+0.88}$ & $60.17_{-1.49}^{+1.26}$ & $2.11_{-1.82}^{+7.02}$ & $48.84_{-2.43}^{+1.29}$ & $-2.00_{-1.00}^{+0.35}$ & $0.31_{-5.69}^{+3.15}$ \\

~ & ~ & ~ & ~ & ~ & ~ & ~ & ~ & ~ & ~\\

\hline
\end{tabular}
\end{center}
\label{tab: classtab}
\end{table*}

\subsection{GRBs classes from cladistics}

 The cladistic analysis is performed on the 213 GRBs described by the seven parameters: $\tau_{90}$, $\log{E_{iso}}$, $\log{L_{iso}}$, $HR$, $\log{L_X}$, $\gamma$ and $\log{{\cal{N}}_{HI}}$. This
yields 9185 equally parsimonious trees. From these, we build a majority-rule consensus tree which provides a summary of all the tree structures. It appears that most nodes are found in all trees, the other ones in more than 80\% of them. This gives an indication that the structure of the majority-rule consensus tree can be trusted.

The classes are then defined on this consensus tree. For this purpose, we select all the objects that belong to obvious sub-branches. Some groups could be further sub-divided, but we have chosen to keep as many objects as possible within each class in order to maintain enough statistical significance (see Table~\ref{tab:contingency}). We thus define nine groups. Ten objects are isolated on the tree and thus cannot define or be assigned to a class.

There is no rigorous statistical way to assign a confidence level to the identified groups. The two usual quantitative indicators are not useful in the present case: bootstrapping (resampling with replacement) of parameters cannot be used because of too few parameters, and the Bremer support (number of supplementary steps of the closest less parsimonious tree for which the node disappears) is difficult to interpret when the number of steps is large (1380 here).

We have also analysed the sub-sample made of the 90 GRBs with an available value for $E_{pk}$ using the same seven parameters. The resulting consensus tree is less robust (a significant number of nodes are found in less than 70\% of the 10 000 equally parsimonious trees), but the structure is in good agreement with the full sample consensus tree. This is a hint that the classes do not depend too much on the sample.  In any case, the best way to assess the significance of the tree structure and the derived classification is to analyse how much the groups are supported by the parameters, which means to see how plausible the partitioning is from the physical point of view. This is described throughout the rest of this paper.

Nevertheless, we can add an independent check that the tree structure from which the groups have been defined does not occur by chance. It is always a good practice in clustering to compare with an independent method. We have performed a k-medoids analysis \citep{kmedoids1987,kmedoids}, which is very different from cladistics: it is a partitioning method based on the pairwise distances, very similar to the better known k-means \citep{kmeans1967,kmeans2010}. This difference generally precludes a full agreement between the groups.

In the k-medoids method like for all partitioning techniques, the number of groups must be given. We have found that for nine groups, three of them are void or nearly so. We present in Table~\ref{tab:contingency} the contingency table between the six groups from k-medoids as compared to the nine classes from cladistics. Cladistics groups 1, 2, 3 and 9 are clearly retrieved by k-medoids, the other ones being split generally into two groups. Even though the agreement is not one to one, as expected, this comparison is very satisfactory. More importantly, we have performed the same work as in the following of this paper with the six groups of k-medoids, and find that the main conclusions are identical.

\begin{figure*}
\centering
\subfigure{\includegraphics[width=3.50cm]{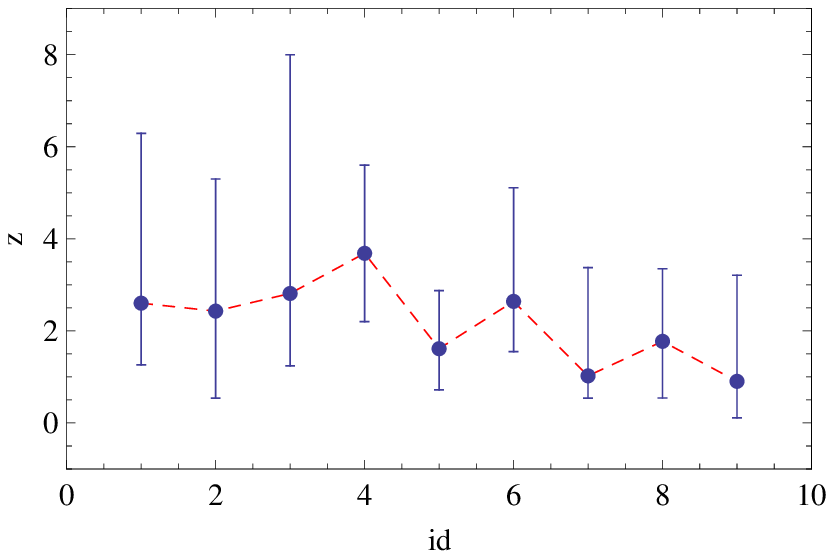}} \goodgap
\subfigure{\includegraphics[width=3.50cm]{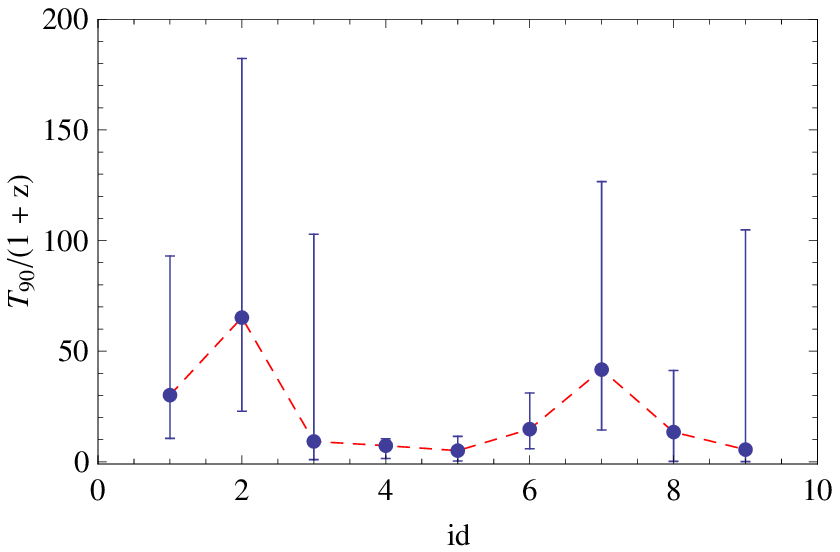}} \goodgap
\subfigure{\includegraphics[width=3.50cm]{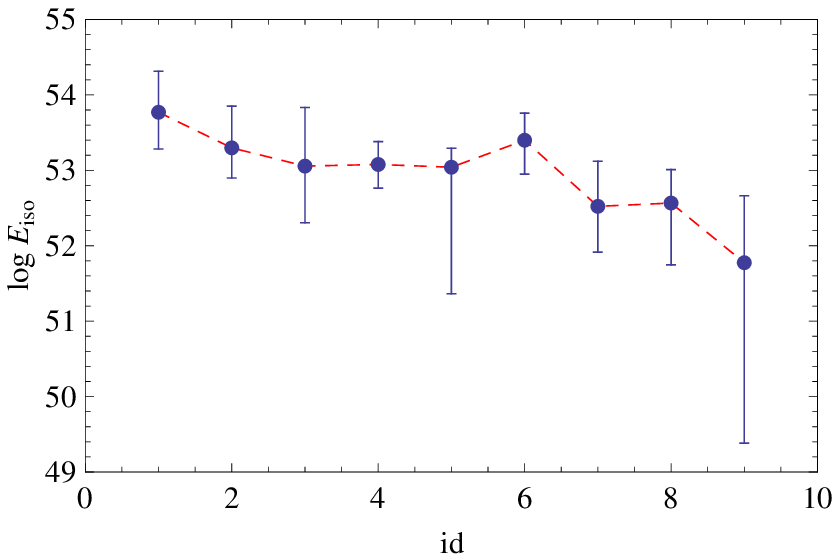}} \goodgap
\subfigure{\includegraphics[width=3.50cm]{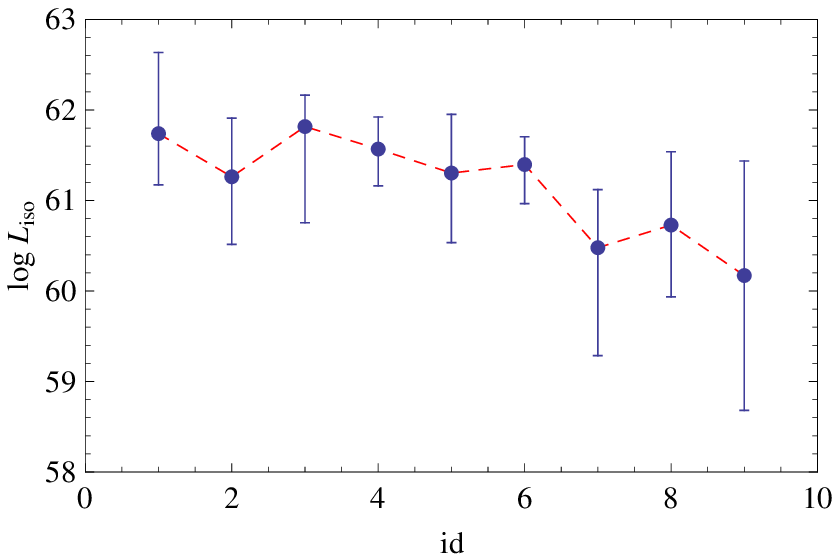}} \goodgap \\
\subfigure{\includegraphics[width=3.50cm]{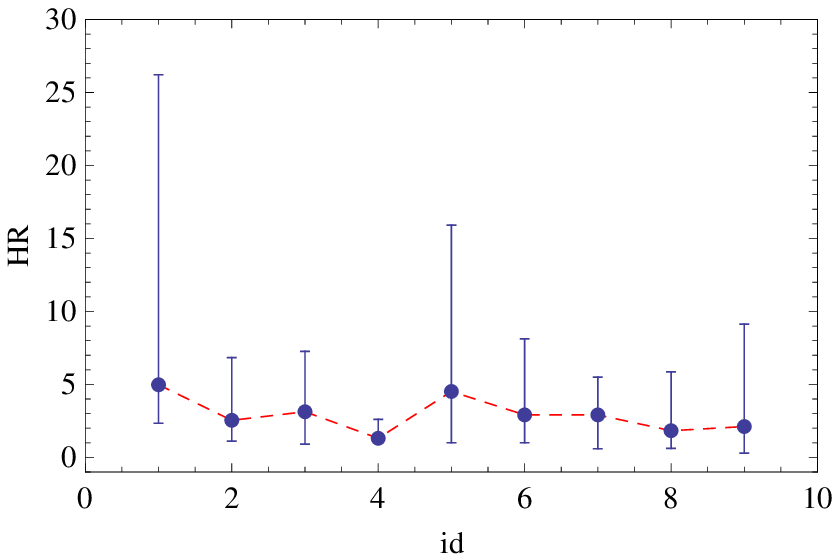}} \goodgap
\subfigure{\includegraphics[width=3.50cm]{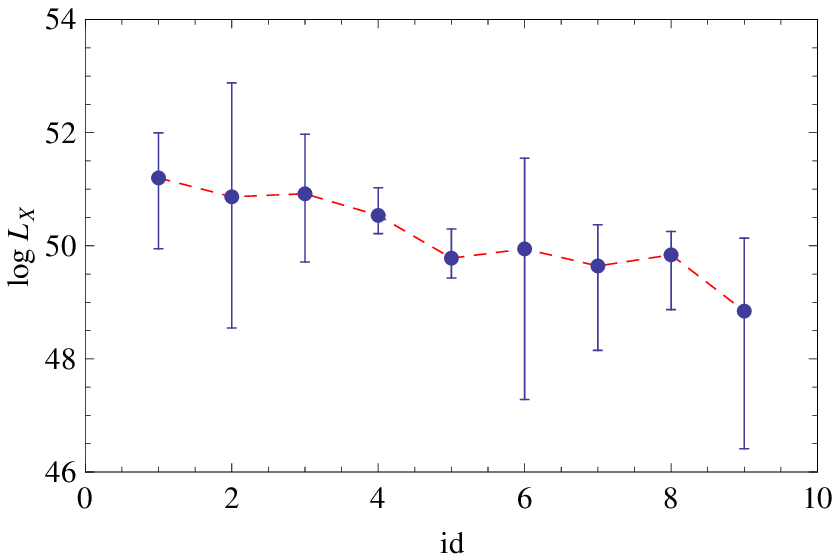}} \goodgap
\subfigure{\includegraphics[width=3.50cm]{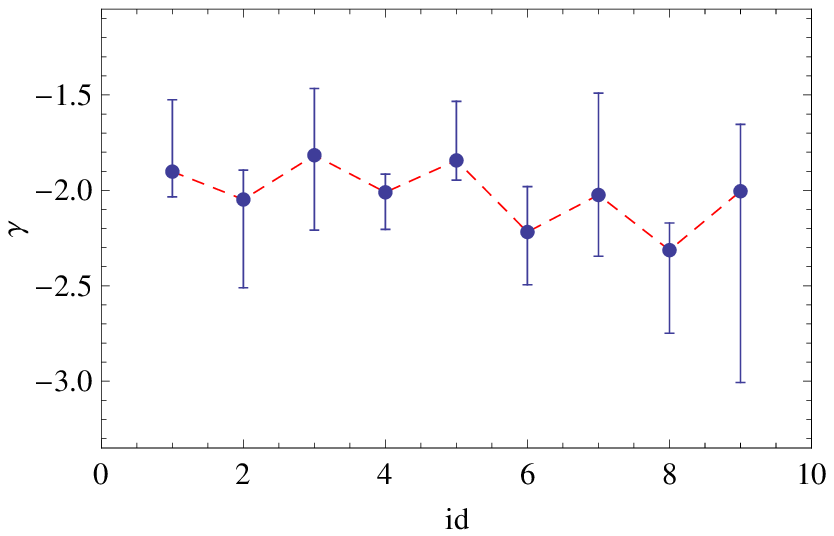}} \goodgap
\subfigure{\includegraphics[width=3.50cm]{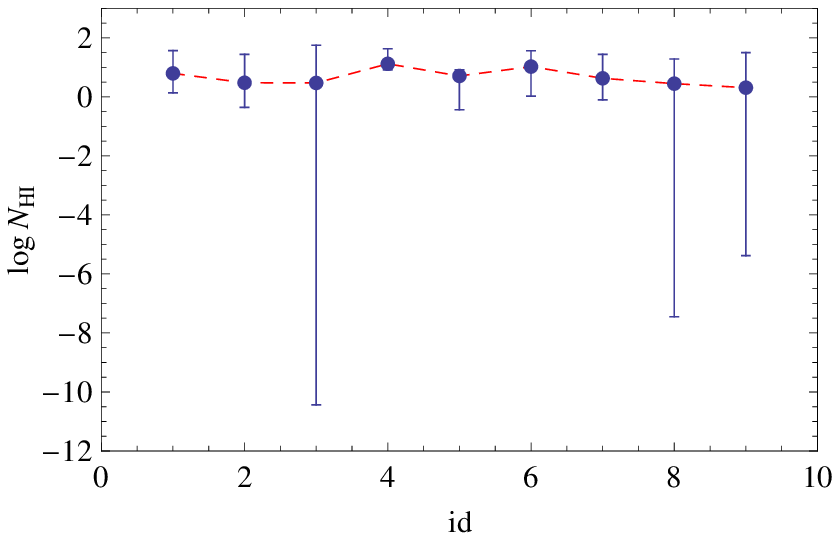}} \goodgap \\
\caption{Median and $90\%$\,CL values of GRB parameters vs the class id. A dashed line is shown to highlight qualitative trends.}
\label{fig: classfig}
\end{figure*}

\begin{figure*}
\centering
\subfigure{\includegraphics[width=7.50cm]{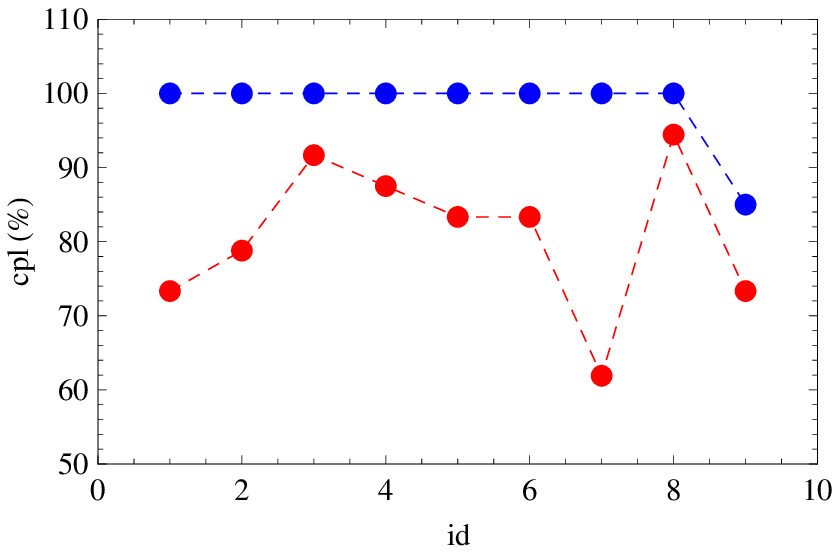}} \goodgap
\subfigure{\includegraphics[width=7.50cm]{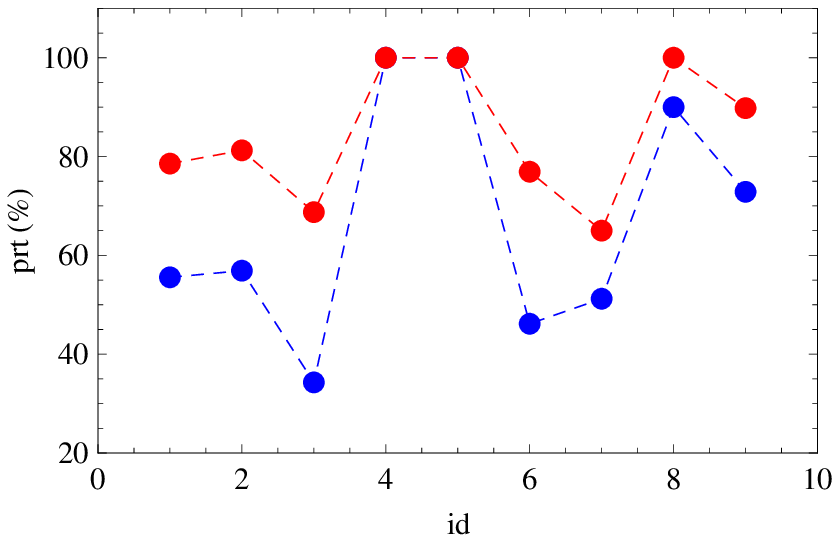}} \goodgap \\
\caption{Completeness (left) and purity (right) for the selection procedure based on median values (blue) and the ${\cal{D}}$ parameter (red).}
\label{fig: cplprtfig}
\end{figure*}

\subsection{Statistical properties}

Having check the robustness of the cladistics analysis, it is worth looking at the statistical properties of the nine identified GRBs classes in order to gain some hints on which are the main parameters differentiating objects belonging to different groups. Although we can not draw any definitive conclusion on the physical mechanism leading to the division in classes, such an analysis may provide a guidance for testing proposed scenarios.

Table\,\ref{tab: classtab} summarizes median and $90\%$ confidence ranges for the redshift $z$ and the input parameters used for the cladistics analysis. Note that, since the distributions are typically strongly asymmetric with very long tails towards one end (partially becuase of poor statistics in some classes), we have preferred to be conservative showing the $90\%$\,CL instead of the more conservative $68\%$ ones to avoid severely underestimating the tails of the histograms. A more user\,-\,friendly look at these same results is provided by Fig.\,\ref{fig: classfig} where the same quantities are plotted vs the class id.

\begin{figure*}
\centering
\subfigure{\includegraphics[width=5.25cm]{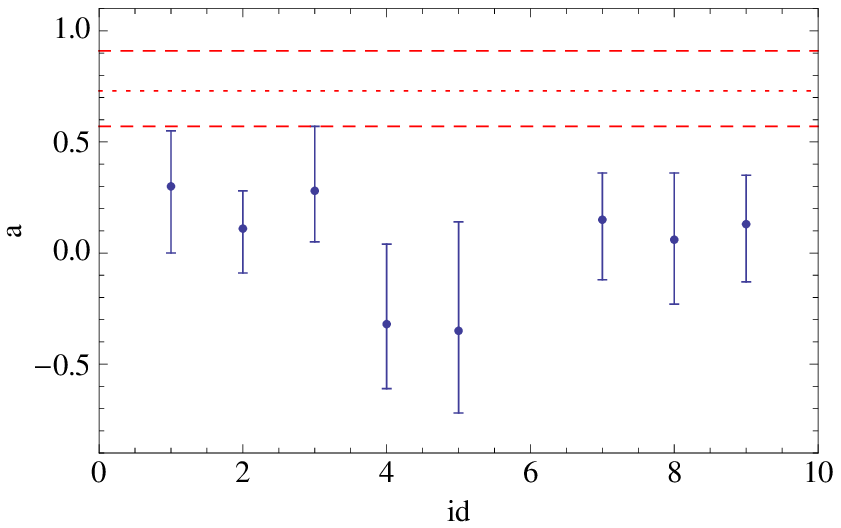}} \goodgap
\subfigure{\includegraphics[width=5.25cm]{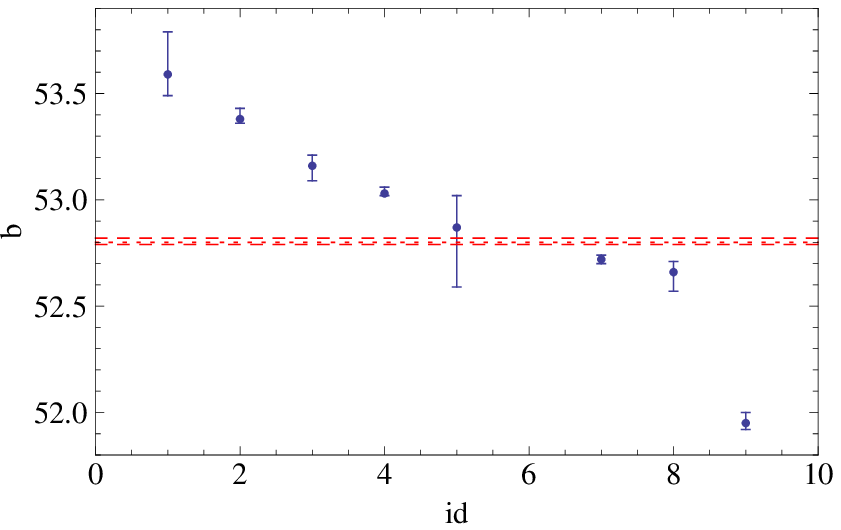}} \goodgap
\subfigure{\includegraphics[width=5.25cm]{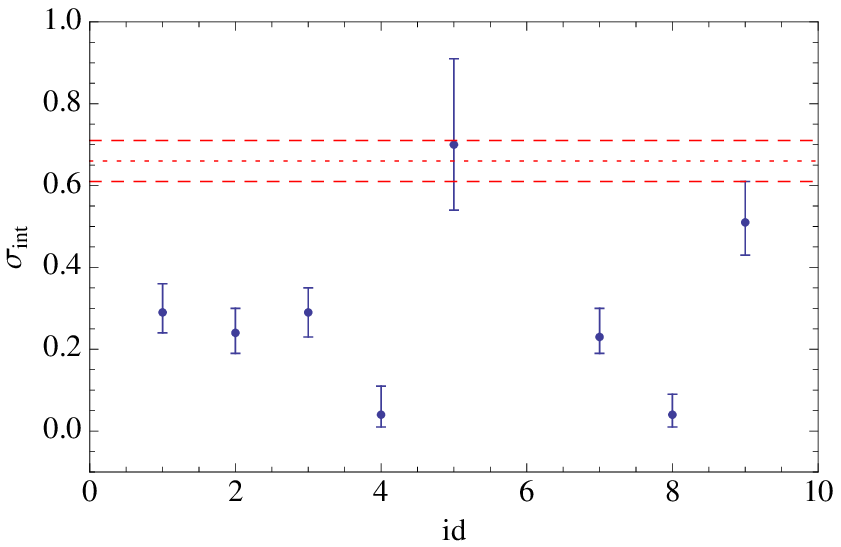}} \goodgap
\caption{Slope $a$, zeropoint $b$ and intrinsic scatter $\sigma_{int}$ of the $E_{pk}$\,-\,$E_{iso}$ correlation vs the class id. Red dotted and dashed lines denote the median and $68\%$\,CL for the fit to the full sample.}
\label{fig: epeisoplot}
\end{figure*}

Although the wide confidence ranges prevent us from drawing definitive conclusions, the trends in Fig.\,\ref{fig: classfig} allow us to infer some interesting qualitative lessons. First, we note that the redshift distributions of the different classes are roughly consistent with each other. Actually, a gently decreasing trend could be inferred from the higher id classes having a lower median redshift. However, the fourth upper panel in Fig.\,\ref{fig: classfig} shows that these bins are populated by the less luminous GRBs which can only be observed up to smaller $z$. It is therefore likely that the smaller median $z$ observed for the higher id classes is the result of selection effects at work. Should this be indeed the case (which could only be verified based on suitable simulated samples), we could argue that the physical mechanism leading to different GRBs classes is redshift independent.

It is interesting to note that $(\log{E_{iso}}, \log{L_{iso}}, \log{L_X})$ clearly correlates with the class id so that GRBs in the higher id classes are the most energetic, luminous and have larger X\,-\,ray luminosity. A weaker similar trend is also found for the hardness ratio HR and the X\,-\,ray spectral slope $\gamma$, while the rest frame time duration $\tau_{90}$ and the column density $\log{{\cal{N}}_{HI}}$ do not show any significant trend. One could naively conclude that these latter quantities do not play a significant role in assigning GRBs to its cladistics class. Actually, this is not the case. Indeed, the cladistics analysis considers the location of GRBs in the full 7D dimensional space looking for homogenous regions which are likely determined by some (hitherto unknown) evolutionary phenomenon. Reducing the parameter space from 7 to 5 dimensions changes the clustering properties of the GRBs thus altering the final classification. Therefore, one must not read the plots in Fig.\,\ref{fig:
classfig} in the reverse way, i.e., one can not infer from the lack of a trend for a parameter $p$ that $p$ is meaningless.

Although quite efficient in finding homogenous GRBs sample, we are nevertheless aware that a cladistics approach is not immediate to apply. We have therefore investigated whether it is possible to assign GRBs to one of the nine classes identified above according to its parameters without performing a full cladistics analysis. To this end, for each GRB, we consider its $(\tau_{90}, \log{E_{iso}}, \log{L_{iso}}, HR, \log{L_X}, \gamma, \log{{\cal{N}}_{HI}})$ values and attach a label $\ell = 1, \ldots, 9$ to it if these quantities enter the corresponding $90\%$ confidence ranges of the $\ell$\,-\,th class. To quantify the efficiency of this selection procedure, we estimate the {\it completeness} (also known as {\it recall}) $cpl$, defined as the fraction of GRBs in a given class which have been correctly assigned to its true class, and the {\it purity} (also referred to as {\it precision}) $prt$, computed as the complement to the number of GRBs which have been incorrectly assigned to that class. Fig.\,\ref{fig:
cplprtfig} shows that this simple selection procedure excellently works in finding all GRBs belonging to a given class with $cpl$ values equal to unity for all classes but the ninth one. However, the price to pay is a high contamination of samples from GRBs not belonging to the chosen class as it is shown by the low $prt$ values.

This is actually not surprising at all. Since the confidence ranges are overlapped, it is possible that the same GRB has multiple labels. In order to discriminate, we then introduce the parameter

\begin{displaymath}
{\cal{D}}(\ell) = \left [ \sum{[{\bf p}_{GRB} - {\bf p}_{med}(\ell)]^2} \right ]^{1/2}
\end{displaymath}
where ${\bf p}_{GRB}$ represent the set of GRB parameters and ${\bf p}_{med}(\ell)$ the median values for the $\ell$\,-\,th class listed in Table\,\ref{tab: classtab}. Note that ${\cal{D}}$ is simply the distance of the given GRB from the median point of the $\ell$\,-\,th class in the 7D parameter space defined by the input quantities used for the cladistics analysis. Fig.\,\ref{fig: cplprtfig} shows that this method allows to significantly increase the purity of the samples, but reduces the $cpl$ values. Deciding which method works best actually depends on which problem one is interested in. Moreover, larger samples are needed to investigate to which extent narrower confidence ranges help in increasing the purity (the completeness) of the first (second) method so that we will postpone this problem to a forthcoming work. As a final remark, we only stress here that, although both proposed procedures are essentially the same as partitioning GRBs according to where they lie in a 7D parameter space, the boundary
of the regions have been obtained using the cladistics analysis which therefore stand out as powerful and efficient to way to partition a 7D space in a limited number of regions.

\section{$E_{pk}$\,-\,$E_{iso}$ and $E_{pk}$\,-\,$L$ correlations}

As already hinted at in the introduction, scaling correlations between GRBs properties have recently attracted a lot of attention because of their potential use as a tool to make this objects standardizeable candles. The significant scatter affecting these empirical 2D laws represents, however, a serious flaw of this approach leading to a GRBs Hubble diagram plagued by large uncertainties on the distance moduli. Lacking a definitive theoretical motivation of these relations, it is still unclear whether the large scatter is related to some unknown physical mechanism or a consequence of forcing GRBs with different properties to follow the same relation. In order to investigate this issue, one should first find a robust method to separate GRBs according to their position in a multi\,-\,parameter space. This is just what we have done here using the cladistics analysis so that it is worthwhile to investigate whether the intrinsic scatter $\sigma_{int}$ and the calibration coefficients $(a, b)$ of a given
correlation depend on the cladistics class input to the fit.

To this end, we consider the $E_{pk}$\,-\,$E_{iso}$ (\cite{A02,A08}) and the $E_{pk}$\,-\,$L$ \citep{S03,Yo04} correlations, where $(E_{iso}, L_{iso}, E_{pk})$ are the already defined isotropic energy and luminosity and the peak energy of the $\nu F_{\nu}$ spectrum\footnote{We have also cross matched our sample to the \cite[hereafter XS10]{XS10} catalog, recently compiled from the authors collecting literature data on GRBs with measured values of the quantities entering popular scaling relations. Unfortunately, when splitted in the nine classes here identified, the number of GRBs turns out to be too small to consider other interesting scaling relations.}.  We fit the data to the log\,-\,linear relation

\begin{equation}
\log{Q} = a \log{R} + b
\label{eq: corrgen}
\end{equation}
with $Q = (E_{iso}, L_{iso})$ and $R = E_{peak}(1 + z)/(300 \ {\rm keV})$. Note that all these quantities are expressed in the GRB rest frame (which motivates the $(1 + z)$ term to scale $E_{peak}$), while the further scaling constant is introduced to minimize the correlation among errors. The calibration coefficients $(a, b)$ and the intrinsic scatter $\sigma_{int}$ are then derived using the Bayesian motivated fitting technique detailed in \cite{Dago05} which takes into account the uncertainties on both variables.

\begin{table}
\caption{Fit results for the $E_{pk}$\,-\,$E_{iso}$ correlation. Columns are as follows\,: 1. class id (set to 0 for the full sample), 2. number of fitted GRBs, 3., 4., 5. median and $68\%$\,CL for $(a, b, \sigma_{int})$. Note that values for class $\ell = 6$ are missing since there is only 1 GRB with measured $E_{pk}$ in this class.}
\begin{center}
\begin{tabular}{ccccc}
\hline \hline
$\ell$ &  ${\cal{N}}$ & $a$ & $b$ & $\sigma_{int}$ \\
\hline
~ & ~ & ~ & ~ & ~  \\

0 & 89 & $0.73_{-0.16}^{+0.18}$ & $52.80_{-0.01}^{+0.02}$ & $0.66_{-0.05}^{+0.05}$ \\

~ & ~ & ~ & ~ & ~  \\

1 & 12 & $0.30_{-0.30}^{+0.25}$ & $53.59_{-0.10}^{+0.20}$ & $0.29_{-0.05}^{+0.07}$ \\

~ & ~ & ~ & ~ & ~  \\

2 & 18 & $0.11_{-0.20}^{+0.17}$ & $53.38_{-0.02}^{+0.05}$ & $0.24_{-0.05}^{+0.06}$ \\

~ & ~ & ~ & ~ & ~  \\

3 & 13 & $0.28_{-0.23}^{+0.29}$ & $53.16_{-0.07}^{+0.05}$ & $0.29_{-0.06}^{+0.06}$ \\

~ & ~ & ~ & ~ & ~  \\

4 & 6 & $-0.32_{-0.29}^{+0.36}$ & $53.03_{-0.01}^{+0.03}$ & $0.04_{-0.03}^{+0.07}$ \\

~ & ~ & ~ & ~ & ~  \\

5 & 7 & $-0.35_{-0.37}^{+0.49}$ & $52.87_{-0.28}^{+0.15}$ & $0.70_{-0.16}^{+0.21}$ \\

~ & ~ & ~ & ~ & ~  \\

7 & 8 & $0.15_{-0.27}^{+0.21}$ & $52.72_{-0.02}^{+0.02}$ & $0.23_{-0.04}^{+0.07}$ \\

~ & ~ & ~ & ~ & ~  \\

8 & 3 & $0.06_{-0.29}^{+0.30}$ & $52.66_{-0.09}^{+0.05}$ & $0.04_{-0.03}^{+0.05}$ \\

~ & ~ & ~ & ~ & ~  \\

9 & 21 & $0.13_{-0.26}^{+0.22}$ & $51.95_{-0.03}^{+0.05}$ & $0.51_{-0.08}^{+0.10}$ \\

~ & ~ & ~ & ~ & ~  \\

\hline \hline
\end{tabular}
\end{center}
\label{tab: fitepeiso}
\end{table}

\begin{figure*}
\centering
\subfigure{\includegraphics[width=7.50cm]{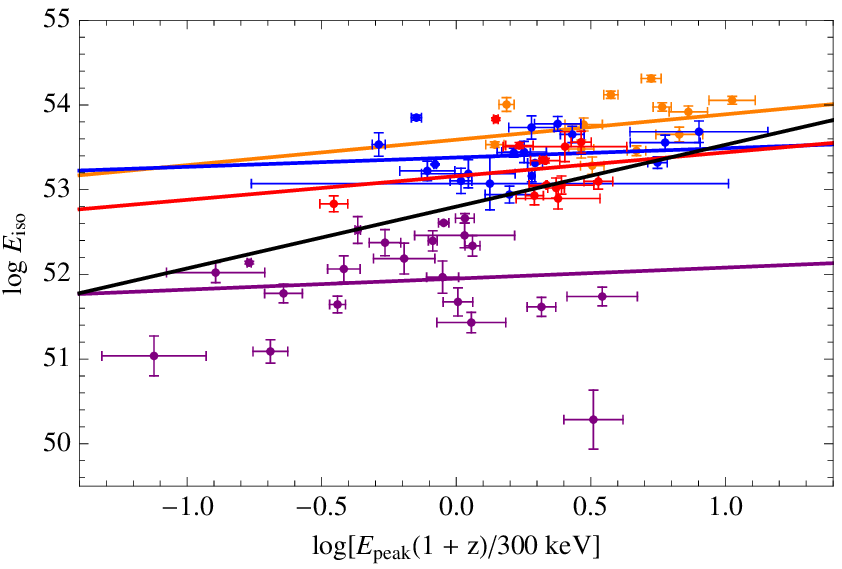}} \goodgap
\subfigure{\includegraphics[width=7.50cm]{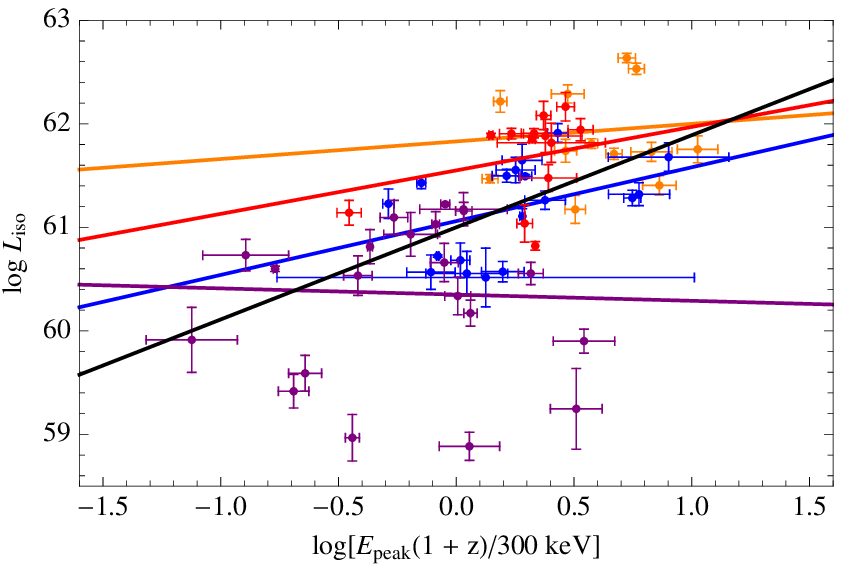}} \goodgap \\
\caption{Median fit superimposed to the data for GRBs classes $\ell = 1$ (orange), $2$ (blue), $3$ (red), $9$ (purple) and for the full sample (black). Left (right) panel refers to the $E_{pk}$\,-\,$E_{iso}$ ($E_{pk}$\,-\,$L_{iso}$) correlation.}
\label{fig: medfitfigs}
\end{figure*}

\begin{figure*}
\centering
\subfigure{\includegraphics[width=5.25cm]{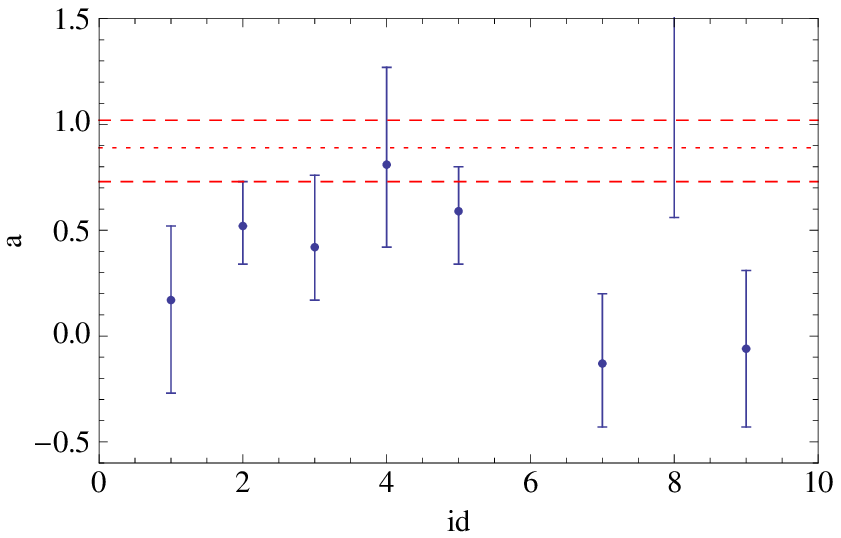}} \goodgap
\subfigure{\includegraphics[width=5.25cm]{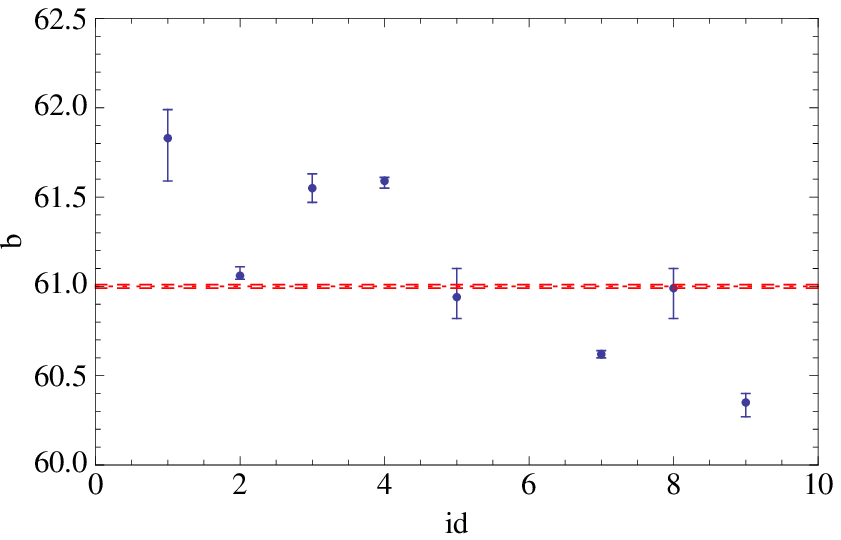}} \goodgap
\subfigure{\includegraphics[width=5.25cm]{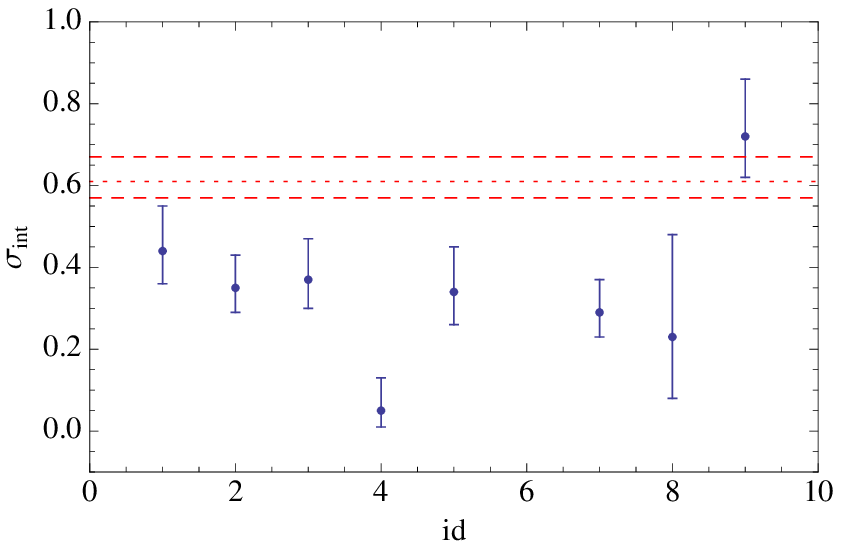}} \goodgap
\caption{Same as Fig.\,\ref{fig: epeisoplot} but for the $E_{pk}$\,-\,$L_{iso}$ correlation.}
\label{fig: eplisoplot}
\end{figure*}

\subsection{Results for the $E_{pk}$\,-\,$E_{iso}$ correlation}

As a first case, we consider the relation between the peak energy $E_{pk}$ and the isotropically emitted one $E_{iso}$ summarizing in Table\,\ref{tab: fitepeiso} the constraints (median and $68\%$ CL) on the calibration coefficients $(a, b)$ and the intrinsic scatter $\sigma_{int}$. Fig.\,\ref{fig: epeisoplot} plots these quantities vs the class id in order to more easily compare them each other and with the results for the fit to the full sample. Note also that we do not have constraints for class $\ell = 6$ since only 1 GRB in this class has a measured $E_{pk}$ value.

Although the small number of GRBs in each subsample leads to large confidence ranges for $(a, b, \sigma_{int})$, some interesting lessons can nevertheless be drawn from Table\,\ref{tab: fitepeiso} and Fig.\,\ref{fig: epeisoplot}. For convenience, let us denote, hereafter, $p_0$ and $p_{\ell}$ the values of the parameter $p$ from the fit to the full sample and to the class $\ell$ subsample, respectively. First, we note that, but for one case, $p_{\ell}$ is not consitent at $68\%$ with $p_0$. In particular, both the slope and the intrinsic scatter turn out to be overestimated from fitting the full sample. On the contrary, $(a, \sigma_{int})$ are roughly consistent from one class to another, while the decreasing trend of $b$ with $\ell$ is a consequence of the larger $\ell$ classes being populated by increasingly less energetic and luminous GRBs as already found above.

The consitency of the slope $a$ and the decrease of $b$ with $\ell$ helps us to understand why $(a, \sigma_{int})$ are so severely overestimated by the fit to the full sample. Indeed, since the fit to the full sample sets a common zeropoint for the relation, the only way to get smaller $E_{iso}$ values for the high $\ell$ GRBs is to increase the slope $a$. This can also be understood looking at Fig.\,\ref{fig: medfitfigs} where we superimpose the median fits (i.e., those obtained by setting the fit parameters to their median values) to the data for classes $\ell = 1, 2, 3, 9$ since they are the more populated ones and the more robust due to the agreement with the results of the k\,-\,medoids analysis. The fit to the full sample roughly coincides with the bisector line of the region delimited by the median fit to the single subsamples thus leading to overstimating $a$. As a consequence, it is not surprising that the scatter is larger since we are trying to mimic different relations with a single one.

Although larger samples are needed to confirm these preliminary results, the present analysis highlights the importance of partitioning GRBs into homoegenous classes before inferring constraints on both the slope and the scatter of the $E_{pk}$\,-\,$E_{iso}$ correlation. Which are the consequences of a biased estimate of $(a, b, \sigma_{int})$ will be addressed later.

\subsection{Results for the $E_{pk}$\,-\,$L$ correlation}

Table\,\ref{tab: fitepliso} and Fig.\,\ref{fig: eplisoplot} shows the results for the $E_{pk}$\,-\,$L$ correlation fitted to the different subsamples and the full sample. As a general remark, we find qualitatively consistent conclusions with the analysis of the $E_{pk}$\,-\,$E_{iso}$ correlation. Indeed, the slope $a$ and the intrinsic scatter $\sigma_{int}$ are typically overestimated by fitting the full sample, while the zeropoint $b$ again decreases with the class id $\ell$ as a consequence of the smaller median luminosity of the higher $\ell$ GRBs. Moreover, the right panel in Fig.\,\ref{fig: medfitfigs} shows that the fit to the full sample roughly follows the bisector line of the region delimited by the fits to the single subsamples.

However, the shallower slope and the larger scatter of the $E_{pk}$\,-\,$L_{iso}$ correlation weaken the constraints on $(a, b, \sigma_{int})$ thus making less robust the above conclusions. We nevertheless note that, for class $\ell = 9$, there seems to be no correlation at all so that forcing these GRBs to follow the same relation as the other ones (which is what one is implicitly assuming when fitting the full sample) has the unpleasant consequence of biasing high the intrinsic scatter.

\begin{table}
\caption{Same as Table\,\ref{tab: fitepeiso} but for the $E_{pk}$\,-\,$L_{iso}$ correlation.}
\begin{center}
\begin{tabular}{ccccc}
\hline \hline
$\ell$ &  ${\cal{N}}$ & $a$ & $b$ & $\sigma_{int}$ \\
\hline
~ & ~ & ~ & ~ & ~  \\

0 & 89 & $0.89_{-0.16}^{+0.13}$ & $61.00_{-0.01}^{+0.01}$ & $0.61_{-0.04}^{+0.06}$ \\

~ & ~ & ~ & ~ & ~  \\

1 & 12 & $0.17_{-0.44}^{+0.35}$ & $61.83_{-0.24}^{+0.16}$ & $0.44_{-0.08}^{+0.11}$ \\

~ & ~ & ~ & ~ & ~  \\

2 & 18 & $0.52_{-0.18}^{+0.21}$ & $61.06_{-0.02}^{+0.05}$ & $0.35_{-0.06}^{+0.08}$ \\

~ & ~ & ~ & ~ & ~  \\

3 & 13 & $0.42_{-0.25}^{+0.34}$ & $61.55_{-0.08}^{-0.08}$ & $0.37_{-0.07}^{+0.10}$ \\

~ & ~ & ~ & ~ & ~  \\

4 & 6 & $0.81_{-0.39}^{+0.46}$ & $61.59_{-0.04}^{+0.02}$ & $0.05_{-0.04}^{+0.08}$ \\

~ & ~ & ~ & ~ & ~  \\

5 & 7 & $0.59_{-0.25}^{+0.21}$ & $60.94_{-0.12}^{+0.16}$ & $0.34_{-0.08}^{+0.11}$ \\

~ & ~ & ~ & ~ & ~  \\

7 & 8 & $-0.13_{-0.30}^{+0.33}$ & $60.62_{-0.02}^{+0.02}$ & $0.29_{-0.06}^{+0.08}$ \\

~ & ~ & ~ & ~ & ~  \\

8 & 3 & $1.77_{-1.21}^{+0.53}$ & $60.99_{-0.17}^{+0.11}$ & $0.23_{-0.15}^{+0.25}$ \\

~ & ~ & ~ & ~ & ~  \\

9 & 21 & $-0.06_{-0.37}^{+0.37}$ & $60.35_{-0.08}^{+0.05}$ & $0.72_{-0.10}^{+0.14}$ \\

~ & ~ & ~ & ~ & ~  \\

\hline \hline
\end{tabular}
\end{center}
\label{tab: fitepliso}
\end{table}

\subsection{Impact on distance estimate}

The high interest in GRBs scaling relations is motivated by the possibility to trace the luminosity distance over a redshift range extending deep into the matter dominated era. As an example, let us consider here the $E_{pk}$\,-\,$E_{iso}$ correlation since it is affected by a smaller scatter. Inverting Eq.(\ref{eq: defeiso}), one trivially gets\,:

\begin{equation}
d_L(z) = \left [ \frac{E_{iso} (1 + z)}{4 \pi S_{bolo}} \right ]^{1/2}
\label{eq: dleiso}
\end{equation}
so that one can estimate $d_L(z)$ from the observable GRB properties (namely, the fluence $S_{obs}$ and the spectral parameters) provided $E_{iso}$ is inferred from the $E_{pk}$\,-\,$E_{iso}$ correlation. Should this quantity be biased because of the use of incorrect calibration parameters $(a, b, \sigma_{int})$, the distance $d_L(z)$ will be biased too. It is only a matter of algebra to show that $D_{all}/D_{bin} = E_{iso}^{all}/E_{iso}^{bin}$ where $(D_i, E_{iso}^{i})$ are the distance and the estimated energy of the GRB at redshift $z$ as estimated from the fit to the full sample ($i = all)$ and to the subsample the GRB belongs to $(i = bin$).

Actually, for a given $E_{pk}$, the correlation predicts that the GRB isotropic energy $\log{E_{iso}}$ follows a normal distribution centred on $\langle \log{E_{iso}} \rangle = a \log{E_{pk}} + b$ and with variance set by the intrinsic scatter $\sigma_{int}$. In order to take care of the uncertainties on the $(a, b, \sigma_{int})$ parameters, we estimate $E_{iso}^{i}$ as follows. For each point $(a, b, \sigma_{int})$ along the merged chain found during the likelihood analysis, we compute

\begin{displaymath}
\langle E_{iso} \rangle \propto \int{E_{iso} {\cal{G}}(\langle \varepsilon \rangle, \sigma_{int}) d\varepsilon}
\end{displaymath}
with $\varepsilon = \log{E_{iso}}$, $\langle \varepsilon \rangle = a \log{E_{pk}} + b$ and ${\cal{G}}(x, \sigma)$ a Gaussian function with mean and variance given by $(x, \sigma)$, respectively. Evaluating this quantity for all the points in the chain gives us a histogram which we use to finally estimate the median and $68\%$\,CL of $E_{iso}$. We then take the values thus obtained using the chains for the fit to the full sample and the separate classes subsamples to finally get $D_{all}/D_{bin}$.

\begin{figure}
\centering
\includegraphics[width=7.5cm]{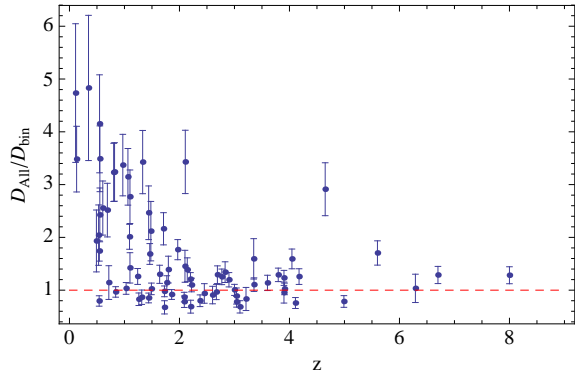}
\caption{Bias on the GRBs distance due to forcing all the objects to follow the same $E_{pk}$\,-\,$E_{iso}$ correlation.}
\label{fig: disttest}
\end{figure}

Considering the full sample, we get (median and $68$ and $95\%$\,CL)\,:

\begin{displaymath}
D_{all}/D_{bin} = 1.29_{-0.42 \ -0.61}^{+1.47 \ +3.54}
\end{displaymath}
showing that the bias on the distance estimate can be quite large and can hardly be neglected. Moreover, as Fig.\,\ref{fig: disttest} shows, $D_{all}/D_{bin}$ is clearly correlated with $z$, the largest biases taking place just over the redshift range $(0, 2)$ where the interplay between dark energy and matter is most important. Although a likelihood and model dependent analyis is needed to quantify the impact on the cosmological parameters, Fig.\,\ref{fig: disttest} is a definitive evidence of the need to preliminary partition GRBs in homogenous classes before using them to infer a cosmologically relevant Hubble diagram.

\section{Conclusions}

The hunt for a distance candle able to extend the Hubble diagram deep into the matter dominated era have had a renewed boost when scaling relations for GRBs became first available. The first successes were, however, soon frustrated by both theoretical (such as the difficulties in understanding their physical motivations) and observational (e.g., the role of selection effects) problems. From the point of view of the Hubble diagram, the most disturbing problem is the significant intrinsic scatter affecting these scaling relations which causes large uncertainties on the inferred distance moduli. Lacking a definitive interpretation of its origin, we have here investigated whether the scatter may be (at least in part) the outcome of forcing a single scaling relation to fit physically different classes of GRBs.

To this end, one must first rely on a robust method to identify which are these classes and to assign a given GRB to the correct class based on its rest frame properties. However, such a task is far to be trivial considering the high dimensionality of the parameter space to be searched for. Cladistics analysis offers a valuable and efficient way to successfully solve this problem. Although it is not possible at the moment to claim which are the physical processes motivating this partitioning of the parameter space, it is worth stressing that all of the GRBs in a given class have undergone the same processes thus representing an homogenous class. They are therefore the ideal inputs to our analysis of how the scaling correlations change with the GRBs subsamples used.

We have then considered the $E_{pk}$\,-\,$E_{iso}$ and $E_{pk}$\,-\,$L_{iso}$ correlations in order to investigate the bias on the calibration parameters $(a, b)$ and the intrinsic scatter $\sigma_{int}$ due to fitting all GRBs together even if they belong to different classes. It turns out that $(a, b, \sigma_{int})$ from the fit to the full sample are remarkably different from those to the separated homogenous classes. These latter fits trace distinct regions of the $(E_{pk}, E_{iso})$ and $(E_{pk}, L_{iso})$ spaces, while the median fit to the full sample approximately trace the diagonal of the region delimited by the single fits. Although a larger statistics is necessary to confirm these preliminary results, it is intriguing to note that the scatter is indeed reduced. Moreover, using the median fit to the full sample to estimate $(E_{iso}, L_{iso})$ for a given GRB instead of the one for its class introduces a non negligible bias on the inferred luminosity distance. Moreover, such a bias is strongly
correlated with the redshift and is larger just over the range most interesting for dark energy studied. Quantifying the impact on the GRBs Hubble diagram and the cosmological parameters determination is outside our aim here, but worth to be investigated in a forthcoming publication.

As a preliminary analysis, one should however perform two complementary tests. First, a larger sample is needed in order to both assess the robustness of the cladistics analysis partitioning and the assignment procedure. Indeed, we have here developed a fast procedure to assign a GRB to its corresponding class based on a multi\,-\,parameter selection criteria based on the analysis of the sample considered here. It is important to validate and improve the completeness and purity of this procedure with an independent sample. This could be quite easy by considering the GRBs observed with other instruments or added to the Swift catalog after our compilation. We can first use our procedure to assign them to one classe and then carry on a full cladistics analysis to check whether the assignment is correct or not. The larger GRBs sample also allows us to improve the determination of the scaling relation parameters $(a, b, \sigma_{int})$ thus offering the possibility to strengthen our conclusion on the importance of
 a preliminary class assignment.

\section*{Acknowledgements}

We warmly acknowledge the unknown referee for his/her constructive report. VFC is also grateful to L. Amati for providing his GRBs compilation in electronic format in advance of publication. VFC is funded by the Italian Space Agency (ASI) through contract Euclid\,-\,IC (I/031/10/0) and the agreement ASI/INAF/I/023/12/0.


\begin{thebibliography}{99}

\bibitem[\protect\citeauthoryear{Amati et al.}{2002}]{A02}
Amati, L., Frontera, F., Tavani, M., in't Zand, J.J.M., Antonelli, A., Costa, E., Feroci, M., Guidorzi, C., et al. 2002, A\&A, 390, 81

\bibitem[\protect\citeauthoryear{Amati et al.}{2008}]{A08}
Amati, L., Guidorzi, C., Frontera, F., Della Valle, M., Finelli, F., Landi, R., Montanari, E. 2008, MNRAS, 391, 577

\bibitem[\protect\citeauthoryear{Amati et al.}{2013}]{A13}
Amati, L., et al. 2013, in preparation

\bibitem[\protect\citeauthoryear{Band et al.}{1993}]{B93}
Band, D., Matteson, J., Ford, L., Schaefer, B.L., Palmer, D. et al. 1993, ApJ, 413, 281

\bibitem[\protect\citeauthoryear{Cardone et al.}{2009}]{CCD09}
Cardone, V.F., Capozziello, S., Dainotti, M.G. 2009, MNRAS, 400, 775

\bibitem[\protect\citeauthoryear{Chattopadhyay et al.}{2007}]{Chat07}
Chattopadhyay, T., Misra, R., Chattopadhyay, A.K., Naskar, M. 2007, ApJ, 667, 1017

\bibitem[\protect\citeauthoryear{D' Agostini}{2005}]{Dago05}
D' Agostini, G. 2005, arXiv\,:\,physics/051182

\bibitem[\protect\citeauthoryear{Felsenstein}{2003}]{Felsenstein2003}
Felsenstein, J. 2003, Inferring Phylogenies (Sinauer Associates, Sunderland,
  Massachusetts)

\bibitem[\protect\citeauthoryear{Fraix-Burnet}{2009}]{DFB09} Fraix-Burnet, D. 2009, In {\it Evolutionary Biology. Concept, Modelization and Application}. Pontarotti, P. (Ed.), Springer Berlin Heidelberg, 363

\bibitem[\protect\citeauthoryear{Fraix-Burnet et~al.}{2012}]{Fraix2012}
Fraix-Burnet, D., Chattopadhyay, T., Chattopadhyay, A.~K., Davoust, E., Thuillard, M. 2012, A\&A, 545, A80

\bibitem[\protect\citeauthoryear{Fraix-Burnet et~al.}{2006{\natexlab{a}}}]{FCD06}
{Fraix-Burnet} D., {Choler} P., {Douzery} E. 2006{\natexlab{a}}, A\&A, 455, 845

\bibitem[\protect\citeauthoryear{Fraix-Burnet et~al.}{2006{\natexlab{b}}}]{jc1}
{Fraix-Burnet} D., {Choler} P., {Douzery} E., {Verhamme} A.
  2006{\natexlab{b}}, {J}ournal of {C}lassification, 23, 31

\bibitem[\protect\citeauthoryear{Fraix-Burnet et~al.}{2009}]{FDC09}
Fraix-Burnet, D., Davoust, E., \& Charbonnel, C. 2009, MNRAS, 398, 1706

\bibitem[\protect\citeauthoryear{Fraix-Burnet et~al.}{2006{\natexlab{c}}}]{jc2}
{Fraix-Burnet} D., {Douzery} E., {Choler} P., {Verhamme} A.
  2006{\natexlab{c}}, {J}ournal of {C}lassification, 23, 57

\bibitem[\protect\citeauthoryear{Gehrels et al.}{2004}]{G04}
Gehrels, N., Chincarini, G., Giommi, P., Mason, K.O., Nousek, J.A., et al. 2004, ApJ, 611, 1005

\bibitem[\protect\citeauthoryear{Ghosh \& Liu}{2010}]{kmeans2010}
Ghosh, J. \& Liu, A. 2010, The Top Ten Algorithms in Data Mining, ed. X.~Wu \&
  V.~Kumar, Chapman \& Hall/CRC Data Mining and Knowledge Discovery Series
  (Taylor \& Francis), 21--36

\bibitem[\protect\citeauthoryear{Kaufman \& Rousseeuw}{1987}]{kmedoids1987}
Kaufman, L. \& Rousseeuw, P. 1987, in Statistical Data Analysis Based on the
  {L1}-Norm and Related Methods, ed. Y.~Dodge (North-Holland), 405--416

\bibitem[\protect\citeauthoryear{MacQueen}{1967}]{kmeans1967}
MacQueen, J.~B. 1967, in Proceedings of 5th Berkeley Symposium on Mathematical
  Statistics and Probability, 281--297

\bibitem[\protect\citeauthoryear{Makarenkov et~al.}{2006}]{Makarenkov2006}
Makarenkov, V., Kevorkov, D., \& Legendre, P. 2006, Applied Mycology and
  Biotechnology, 6 - Bioinformatics

\bibitem[\protect\citeauthoryear{Nixon}{1999}]{ratchet}Nixon, K.~C. 1999, Cladistics, 15, 407

\bibitem[\protect\citeauthoryear{Paciesas et al.}{1999}]{Batse}
Paciesas, W.S., Meegan, C.A., Pendleton, G.N., Briggs, M.S., Kouveliotou, C., et al. 1999, ApJS, 122, 465

\bibitem[\protect\citeauthoryear{Reynolds et~al.}{2006}]{kmedoids}
Reynolds, A., Richards, G., Iglesia, B., \& Rayward-Smith, V. 2006, Journal of
  Mathematical Modelling and Algorithms, 5, 475

\bibitem[\protect\citeauthoryear{Sakamoto et al.}{2011}]{Saka11}
Sakamoto, T., Barthelmy, S.D., Baumgartner, W.H., Cummings, J.R., Fenimore, E.E., et al. 2011, ApJS, 195, 2

\bibitem[\protect\citeauthoryear{Schaefer}{2003}]{S03}
Schaefer, B.E. 2003, ApJ, 583, L67

\bibitem[\protect\citeauthoryear{Schaefer}{2007}]{S07}
Schaefer, B.E. 2007, ApJ, 660, 16

\bibitem[\protect\citeauthoryear{Semple \& Steel}{2003}]{semple2003}
{Semple}, C., {Steel}, M.~A. 2003, Phylogenetics (Oxford University Press)

\bibitem[\protect\citeauthoryear{Swofford}{2003}]{paup}
{Swofford} D.~L., 2003, PAUP*: Phylogenetic Analysis Using Parsimony (*and
  Other Methods), Sinauer Associates, Sunderland, Massachusetts
  (http://paup.csit.fsu.edu/)

\bibitem[\protect\citeauthoryear{Xiao \& Schaefer}{2010}]{XS10}
Xiao, L., Schaefer, B.E., 2010, ApJ, xxx, yyy

\bibitem[\protect\citeauthoryear{Yonekotu et al.}{2004}]{Yo04}
Yonetoku, D., Muramaki, T., Nakamura, T., Yamazaki, R., Inoue, A.K., Ioka, K. 2004, ApJ, 609, 935

\end{thebibliography}
\end{document}